# Evolutionary Algorithm Guided Voxel-Encoding Printing of Functional Hard-Magnetic Soft Active Materials

*Shuai Wu, Craig M. Hamel, H. Jerry Qi, Ruike Zhao*

Shuai Wu, Ruike Zhao
Department of Mechanical and Aerospace Engineering, The Ohio State University, Columbus, OH, 43210, USA
E-mail: zhao.2885@osu.edu
Craig M. Hamel, H. Jerry Qi
The George W. Woodruff School of Mechanical Engineering, Georgia Institute of Technology, Atlanta, GA 30332, USA
E-mail: qih@me.gatech.edu

**Abstract**:

Hard-magnetic soft active materials (hmSAMs) have attracted a great amount of research interests due to their fast-transforming, untethered control, as well as excellent programmability. However, the current direct-ink-write (DIW) printing-based fabrication of hmSAM parts and structures only permits programmable magnetic direction with a constant magnetic density. Also, the existing designs rely on the brute-force approach to generate the assignment of magnetization direction distribution, which can only produce intuitional deformations. These two factors greatly limit the design space and the application potentials of hmSAMs. In this work, we introduce a *voxel-encoding DIW printing* method to program both the magnetic density and direction distributions during the hmSAM printing. The voxel-encoding DIW printing is then integrated with an evolutionary algorithm (EA)-based design strategy to achieve the desired magnetic actuation and motion with complex geometry variations and curvature distributions. With the new EA‐guided voxel-encoding DIW printing technique, we demonstrate the functional hmSAMs that produce complicated shape morphing with desired curvature distributions for advanced applications such as biomimetic motions. These demonstrations indicate that the proposed EA‐guided voxel-encoding DIW printing method can significantly broaden the application potentials of the hmSAMs.

**Introduction**

Hard-magnetic soft active materials (hmSAMs), a functional soft composite that consists of hard-magnetic particles embedded in a soft matrix, have attracted a great amount of research interests due to their fast-transforming, untethered control as well as excellent programmability, promising applications in soft robotics,[1] active metamaterials,[2] morphing devices,[3] and biomedical devices.[4] As shown in **Figure 1**a, the embedded hard-magnetic particles, once being magnetized, can provide micro-torques and deform the soft matrix to mechanically align their magnetic polarities with the externally applied magnetic field $B$. To enable functional actuation of hmSAMs with complex deformation, a well-designed magnetization distribution needs to be encoded into the structure. To increase the programmability and fabrication flexibility of hmSAMs, the additive manufacturing technique of direct-ink-write (DIW) was recently developed, as illustrated in Figure 1b.[2a] Here, by applying a longitudinal magnetic field near the tip of the printing nozzle, the magnetic ink has a magnetization along the printing direction as the magnetic polarities of the embedded particles are aligned with the longitudinal magnetic field. By controlling the printing path, a predesigned distribution of the magnetization direction can be encoded into the printed structure and thus generate shape changes upon the application of an actuation magnetic field (Figure 1c). To predict the magnetic actuation, a theoretical framework was developed and implemented through finite element method (FEM) to describe the magneto-mechanical behavior of hmSAMs.[5] Enabled by this simulation platform, the complex shape change of functional hmSAMs with programmed magnetization direction distribution can be precisely predicted (Figure 1c).

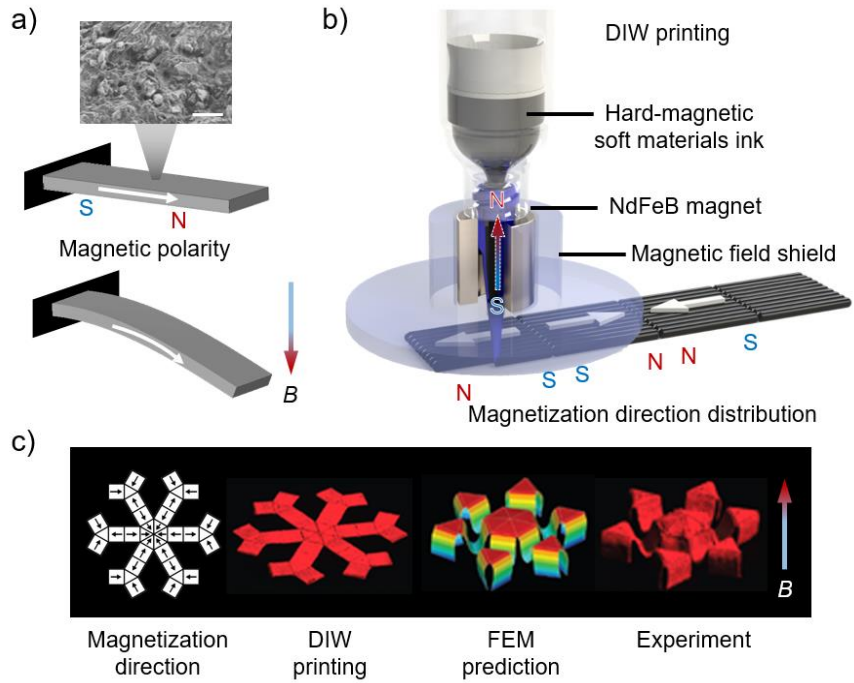

**Figure 1.** Actuation and direct ink writing (DIW) printing of hard-magnetic soft active materials (hmSAMs). a) The actuation mechanism of hmSAM. Scale bar: 15 μm. b) DIW printing of hmSAMs with programmed distribution of magnetization direction. c) A DIW printed sample with the complex shape actuation and the FEM prediction.

Sophisticated functionalities usually rely on shape changes with nonuniform curvatures, or more specifically, with precisely controlled curvature distribution.[6] For example, a sidewinder (**Figure 2**a) can move at a high speed on both hard and granular surfaces. It slithers on a sand dune by reducing slippage at its contact points with the sand and manages its body's agile deformation morphology during winding to provide efficient motion.[7] The morphable long tail of a cheetah plays a significant role while it is hunting prey at a very high speed (~70mph). The tail, which acts like a rudder, morphs itself with a well-designed curvature distribution, allowing the cheetah to quickly change direction and providing counterbalance as it zigzags across grasslands during a chase (Figure 2b).[8] All these efficient motions are achieved by precisely and synergistically controlling the dynamic curvature distribution of the body parts or even the entire body. Although DIW printing of hmSAMs brings the feasibility for the magnetization direction (M-direction) distribution control that allows complicated shape change, the existing designs rely on brute-force approach to generate the assignment of M-direction distributions, which can only produce intuitional deformations. Additionally, current DIW printing of hmSAMs only permits programmable M-direction with a constant magnetization density (M-density), as the magnetic field for alignment is predetermined by the printing setup. However, to achieve complex functional curvature distribution of the deformed shape, both M-density and M-direction need to be rationally programmed during printing, which is impossible via the current DIW printing technique and the brute-force approach for magnetization distribution design due to the complexity and highly nonlinearity of the deformation morphology. Thus, the design space and the potential functionalities and applications of hmSAMs are significantly limited. Note that although it is technically feasible to tune the M-density during printing by applying a varying magnetic field produced by an electromagnetic coil, it is impossible to print an hmSAM filament with a low M-density after printing with a high M-density, as the magnetic polarities of the magnetic particles in the hmSAM ink will be aligned by the higher M-density field and cannot return to the lower M-density state.

In this paper, to facilitate the design of complex actuation of hmSAMs for functional applications, we introduce a *voxel-encoding DIW printing* method to program the *magnetization distribution*, i.e., both the M-density and M-direction distributions, during printing of the hmSAMs. The voxel-encoding DIW printing is then integrated with an evolutionary algorithm (EA)-based design strategy to achieve the desired magnetic actuation and motion with complex geometry variations and curvature distributions.[9] With the new EA-guided voxel-encoding DIW printing technique, we demonstrate the functional hmSAMs that produce complicated shape morphing with desired curvature distributions for advanced application such as biomimetic motions.

**Results**

**Tunable magnetization by voxel-encoding DIW printing of the hmSAMs**

For a printed hmSAM, its M-density and M-direction distributions determine the strength of the driving force and deformation trajectory under an external magnetic field, respectively. As shown in Figure 2c, if the hmSAM beam has an alternating M-direction distribution but a constant M-density in each section, it deforms into a symmetric wavy shape. To achieve functional deformations and dynamic motions, a symmetry-breaking actuation is necessary.[2c] As discussed above, the well-controlled dynamic curvature distribution of a sidewinder snake is the key to achieve the high speed sidewinding motions on a sand surface (Figure 2a). By tuning the M-density and M-direction distributions of the hmSAM beam at the same time, shape morphing with more complicated curvature distributions can be obtained to facilitate functional applications such as biomimetic deformations and motions. Here we introduce a voxel-encoding DIW printing method that can tune the magnetization (both the M-density and M-direction) of the hmSAMs at the voxel-level. In this printing method, each voxel consists of multiple high-aspect-ratio DIW printed hmSAM layers as shown by the schematic graph and an actual DIW printed voxel in Figure 2d. By controlling the printing direction of each layer in the voxel, the magnetization of the entire voxel can be programmed. For example, Figure 2e presents a DIW printed voxel with the number of layers $n = 3$, then the effective magnetization can be encoded with $2n+1 = 7$ variations (represented by the gradient colors) due to the combination of inactive layer "I", left-print layer "L", and right-print layer "R". The magnetization of the three-layered voxel is plotted with respect to the seven different voxel-encoding types with left defined as the positive M-direction shown in Figure 2f. It shows that M-density increases linearly with the number of active hmSAMs layers printed in the same direction. Note that the effective M-density of the voxel is the vector summation of the printed layers, i.e. [LIL] has the same M-density as [LLI]; [LRL] has the same M-density as [LII] (see Supporting Information for more details). By increasing the layers in a DIW voxel, finer M-density tunability can be realized. For example, a four-layered voxel and a five-layered voxel generate 9 and 11 voxel-encoding types, respectively (See the VSM measurement of M-density in Supporting Information).

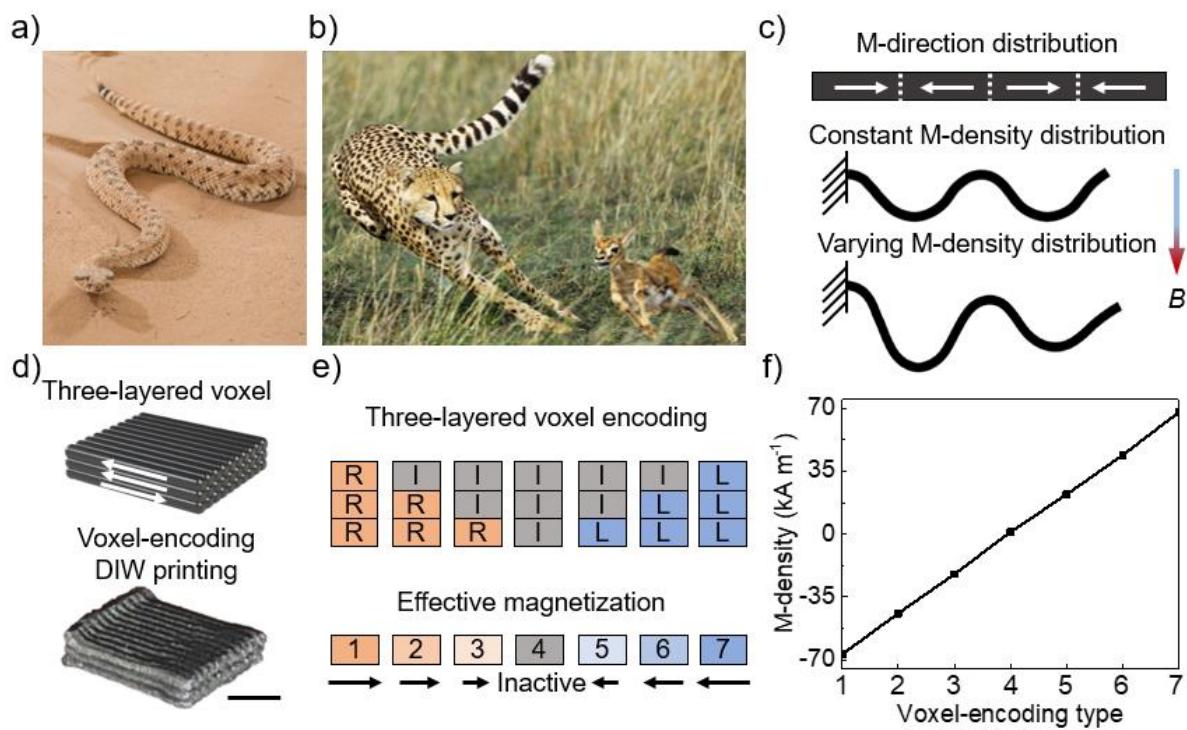

**Figure 2.** Voxel-encoding DIW printing of hmSAMs. a) A sidewinder snake controls its body curvature distribution for the sidewinding motion. b) A cheetah's morphing tail helps balancing and quick direction changing during chasing a prey. c) Actuation schematics of hmSAMs with tunable M-direction and M-density. d) Voxel schematic and a printed voxel by the voxel-encoding DIW printing. Scale bar: 2 mm. e) Three-layered voxel encoding with tunable M-density and M-direction (or magnetization). f) Measured M-density of voxel-encoding DIW printed three-layered voxels.

**An EA-guided strategy to program magnetization distributions**

Utilizing the voxel-encoding DIW printing, high magnetization tunability of hmSAMs is achieved, which can greatly enhance the programmability of hmSAM systems with much broader design space. For a DIW printed beam composed of *m* voxels with *n* layers, the design space is $(2n+1)^m$. For example, for a beam consisting of $m = 10$ voxels with $n = 3$ layers as shown in **Figure 3**a, the design space is $7^{10} = 282,475,249$. Although the voxel-encoding DIW printing provides solution for programming the hmSAMs, the assignment of magnetization in individual voxels to achieve a predetermined shape change is impossible via the brute-force approach due to the large number of variations and high complexity of the voxel setting. Here, we introduce an EA-guided design strategy to address this challenge of programming magnetization distributions for complex curvature distributions and dynamic motions. To achieve a highly autonomous inverse design for magnetization distributions for voxel-encoding DIW printing, the desired beam deformation is fed into the EA-guided design strategy. After iterations of EA-based optimization, the deformation of the structure from the FEM simulation evolves towards the target morphology and curvature distribution. When a certain criterion is reached, the EA process terminates and exports the magnetization distribution, as shown in Figure 3a. As an example of the EA-guided design process (Figure 3b), we consider a beam that is composed of $m = 10$ voxels with $n = 3$ layers in each voxel. As discussed above, individual voxels have seven variations, which are termed as 1, 2, 3, 4, 5, 6, and 7, as genotypes. Individual's genotypes are translated into M-density and M-direction distributions, which are used to construct the FEM models. The deformation from the FEM simulation is used to evaluate the generated magnetization distribution through a fitness function (see Supporting Information for more details). Selection, crossover, and mutation steps of the EA are performed by using an open source evolutionary computation framework to create new magnetization distribution until the desired deformation with a small enough fitness function value or the predetermined maximum number of generations is reached.[10] It should be noted that with the effective magnetization concept, the total number of variables in the model is significantly reduced. If using individual layers' M-direction (inactive, left-print, and right-print) as the design variables, the total design space would be $(3^n)^m$. For the case in Figure 3 with $m = 10$ voxels with $n = 3$ layers in each voxel, it is $3^{30}$, which is ~728,881 times larger than the current one ($7^{10}$) yet offers the same design flexibility. With the integration of the EA-guided design and the voxel-encoding DIW printing of hmSAMs, the magnetization distributions can be rationally programmed, enabling the deformation under the magnetic actuation with desired curvature distributions for predetermined functionalities.

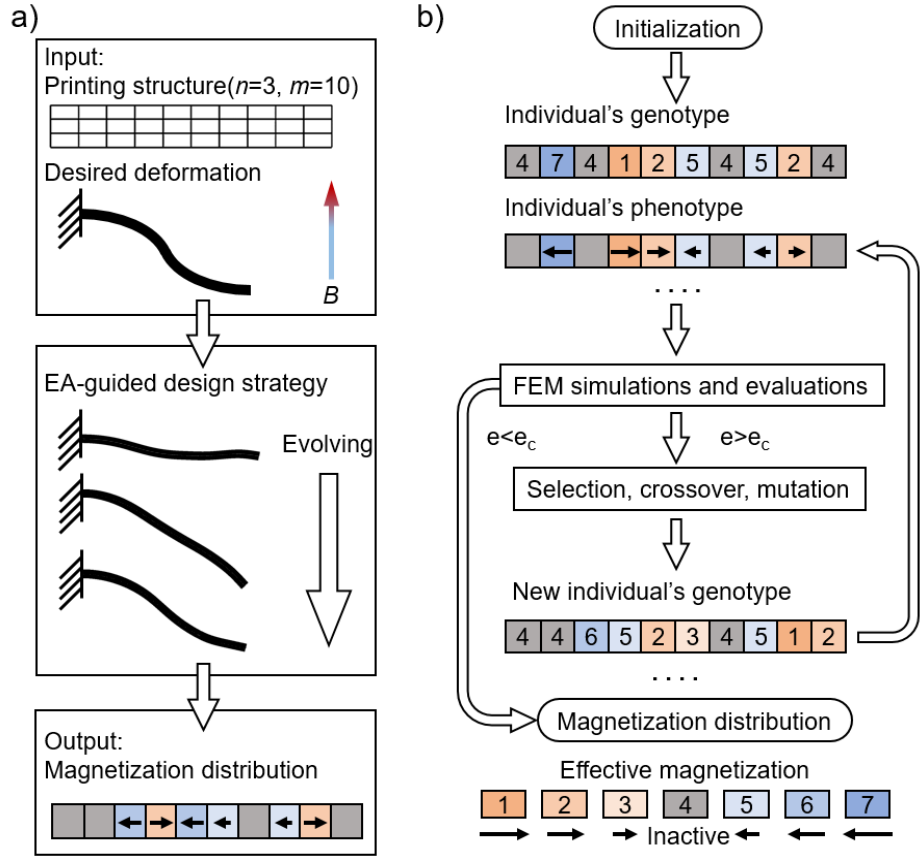

**Figure 3.** EA-guided design strategy for programming magnetization distribution of hmSAMs. a) With desired curvature distribution as the input, the magnetization distribution are generated by EA. b) Flowchart of the EA-guided design strategy.

**Effect of voxel size on the EA-guided design and DIW printing**

Here, we start from using a three-layered ($n = 3$) voxel with 7 voxel-encoding types, which provide fine enough magnetization tunability. Intuitively, for an hmSAM beam with a number of voxels, decreasing the length of the voxels would increase the resolution for deformation. However, a small voxel size affects the actual DIW printing quality due to the accumulated ink cluster at the two edges of the printing filament. Additionally, it slows down the printing speed due to the additional motions of the printing nozzle. To decide the voxel size for high enough printable resolution, 50 mm-long beams with voxels lengths of 1mm, 2 mm, 5 mm, and 10 mm ($m = 50$, 25, 10, and 5, respectively) are evaluated for their printability, magnetic property, mechanical property, and deformation resolution. **Figures 4**a-c shows the EA-calculated magnetization distributions for the three cases (2mm, 5mm, and 10mm) when actuating the beams to a quarter-circle target shape shown in Figure 4d. The M-densities of the three-layered voxel are represented by the gradient colors: dark blue, blue, and light blue for left-direction M-densities of $M$, $2/3M$, and $1/3M$; dark orange, orange, and light orange for right-direction M-densities of $M$, $2/3M$, and $1/3M$, where $M$ is the M-density of a single printed hmSAM filament. (See Supporting Information for the generated magnetization distribution of the hmSAM layers). As an example, Figure 4d shows that, under a 40 mT upward magnetic field, the beam with 2 mm voxels deforms into the desired target shape, with the EA-calculated magnetization distributions. Figure 4e shows the FEM predictions of the magnetic actuations for beams with 2 mm, 5 mm and 10 mm voxels, denoted by the diamonds, triangles and circles, respectively. The solid black curve is the target quarter-circle shape. The comparisons show good agreements between the FEM-predicted deformed shapes and the target shape even for the beam with relatively long voxels (10 mm). Though the fitness function in Figure 4f shows that the beam with 1 mm voxel achieves the best actuation in the FEM simulations and EA calculations, designs with smaller voxel size are much more time-consuming during printing because the motion control of the nozzle constantly requires its moving up and down as it prints. Figure 4f also shows the printing time as a function of the voxel length. To print a three-layered 50 mm by 4.8 mm cantilever beam, the printing time increases 11 times when using 1 mm voxels than using 10 mm ones. Next, the printing quality is taken into consideration by evaluating the appearance, the mechanical and magnetic properties of the DIW voxels. From the printed single-layered 20 mm by 4.8 mm stripes with different voxel length in Figure 4g, the printing quality decreases with the voxel length as the shorter voxel leads to accumulated clusters between adjacent voxels. As a result, the M-density at the clusters is very weak, which causes the significant decrease of M-density for the whole beam when printed using small voxels, as shown in Figure 4h. Here, the ideal M-density is its largest possible value and is measured from an hmSAM sample that is magnetized after solidification represented by the black dashed line in Figure 4h. With 1 mm printing voxel, the measured M-density can only reach 40.8 kA m$^{-1}$, which is only 36.2% of the ideal M-density and greatly limits the range of the tunable M-density. With larger printing voxels, the M-density increases and reaches a plateau with a maximum value of 70.7 kA m$^{-1}$ (62.8% of the ideal M-density). While the voxel size drastically affects the magnetic property, the mechanical property indicates a relatively consistent shear modulus around 300.5 kPa, as shown in Figure 4h. It should be noted that the M-density of 70.7 kA m$^{-1}$ and shear modulus of 300.5 kPa are used in FEM simulations and EA calculations for the target deformation shape for beams with different voxel sizes. Due to the reduced M-density for beam with small printing voxels, the deformation under the same magnetic field decreases. Consequently, Figure 4i indicates that under the same applied magnetic field, the beams with small voxel sizes (1 mm, 2 mm) deform less than the beams with large voxel sizes (5 mm, 10 mm). To reach the same target quarter-circle shape, an increased magnetic field is needed for beams with 1 mm and 2 mm printing voxels (Supporting Information, Video S1). Based on these studies on the voxels, to ensure the DIW printing quality and speed, as well as the EA-guided design's accuracy, 5 mm voxel is chosen for the EA-guided voxel-encoding DIW printing of the hmSAMs in the rest of the paper to achieve the programmable magnetization distribution.

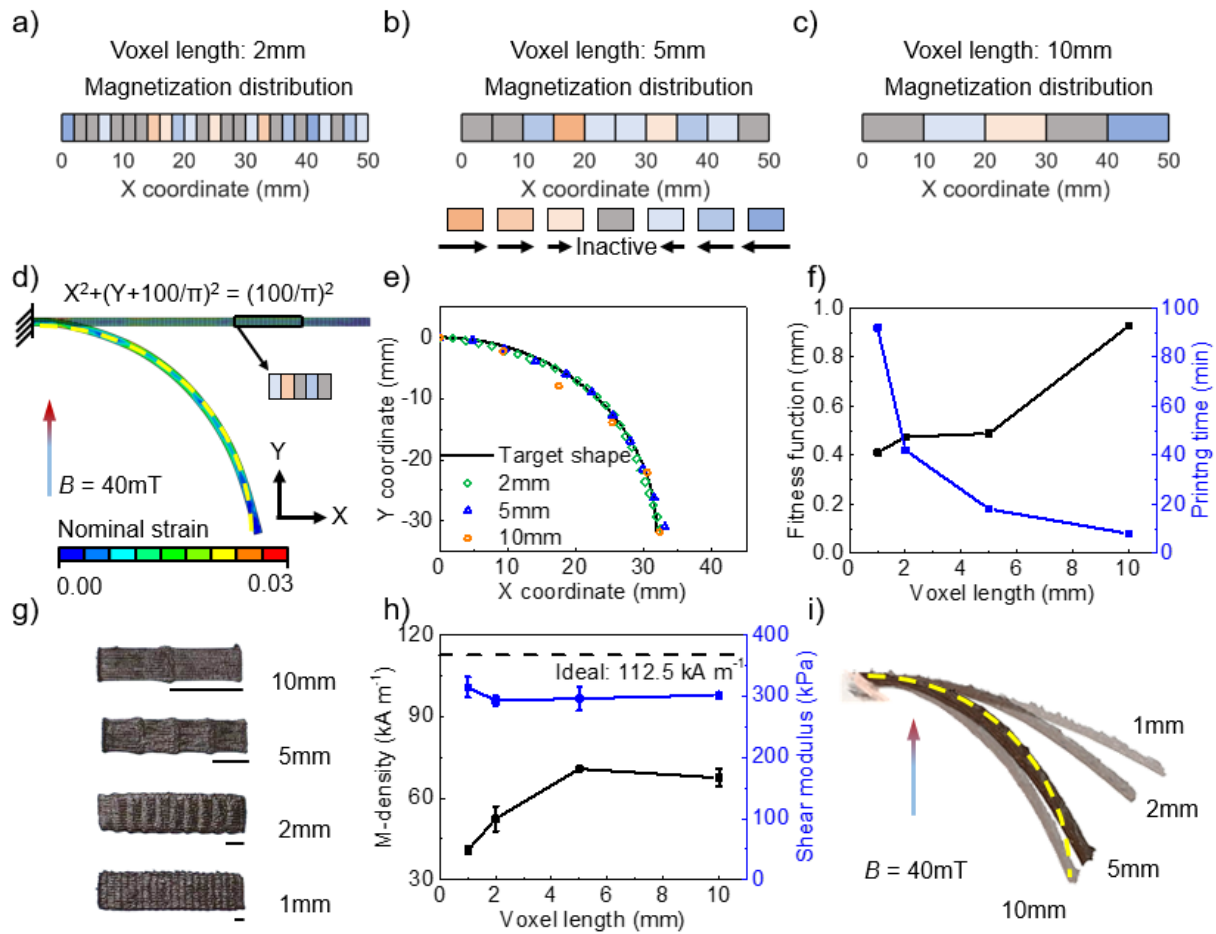

**Figure 4.** Effect of voxel size on the EA-guided design strategy and voxel-encoding DIW printing. a-c) Magnetization distributions with 2 mm, 5 mm and 10 mm three-layered voxels. d) Maximum principal nominal strain distribution of the deformed beam with 2 mm voxel under 40 mT. e) Comparisons between the target shape and the FEM simulations with different voxel lengths. f) Fitness function and DIW printing time with respect to voxel length. g) DIW printed single-layered 20 mm by 4.8 mm samples with 10 mm, 5 mm, 2 mm and 1 mm-long voxels. Scale bars: voxel lengths. h) M-density and shear modulus of DIW printed samples with respect to voxel length. Black dashed line represents the ideal M-density of an hmSAM sample that is magnetized after solidification. i) Magnetic actuation of DIW printed EA-guided beam designs in (a)-(c) under 40 mT.

**EA-guided magnetization distribution for targeted deformations**

Utilizing the EA-guided voxel-encoding DIW printing of the hmSAMs with tunable magnetization distributions, complicated morphing structures, including parabolic, cosine, and half-circle curves, are demonstrated in this section. **Figures 5**a-c show the magnetization distributions of the three targeted deformations from the FEM simulations and EA calculations. The corresponding FEM simulations of the EA-guided design under 20 mT, 30 mT and 35 mT are shown in Figures 5d-f respectively, with the strain contours showing the maximum principal nominal strain distributions of the deformed beams. Figures 5g-i show the comparisons between the target shapes (the solid black curves) and deformed shapes from FEM simulations (the orange circles), indicating good agreements for all three cases. Using the magnetization distributions from the EA-guided design strategy, three different 50 mm-long beams composed of three-layered 5 mm-long voxels with different magnetization distributions are printed. Figures 5j-l and Video S2 show the actuation of these beams under 21 mT, 32 mT and 32 mT magnetic fields, indicating good agreements with the FEM simulations. The highly autonomous and reliable EA-guided design strategy for voxel-encoding DIW printing of hmSAMs with programmable magnetization distributions enables various desired magnetically actuated deformations with predesigned curvature distributions and further expands the application possibilities in reconfigurable and functional structures and devices.

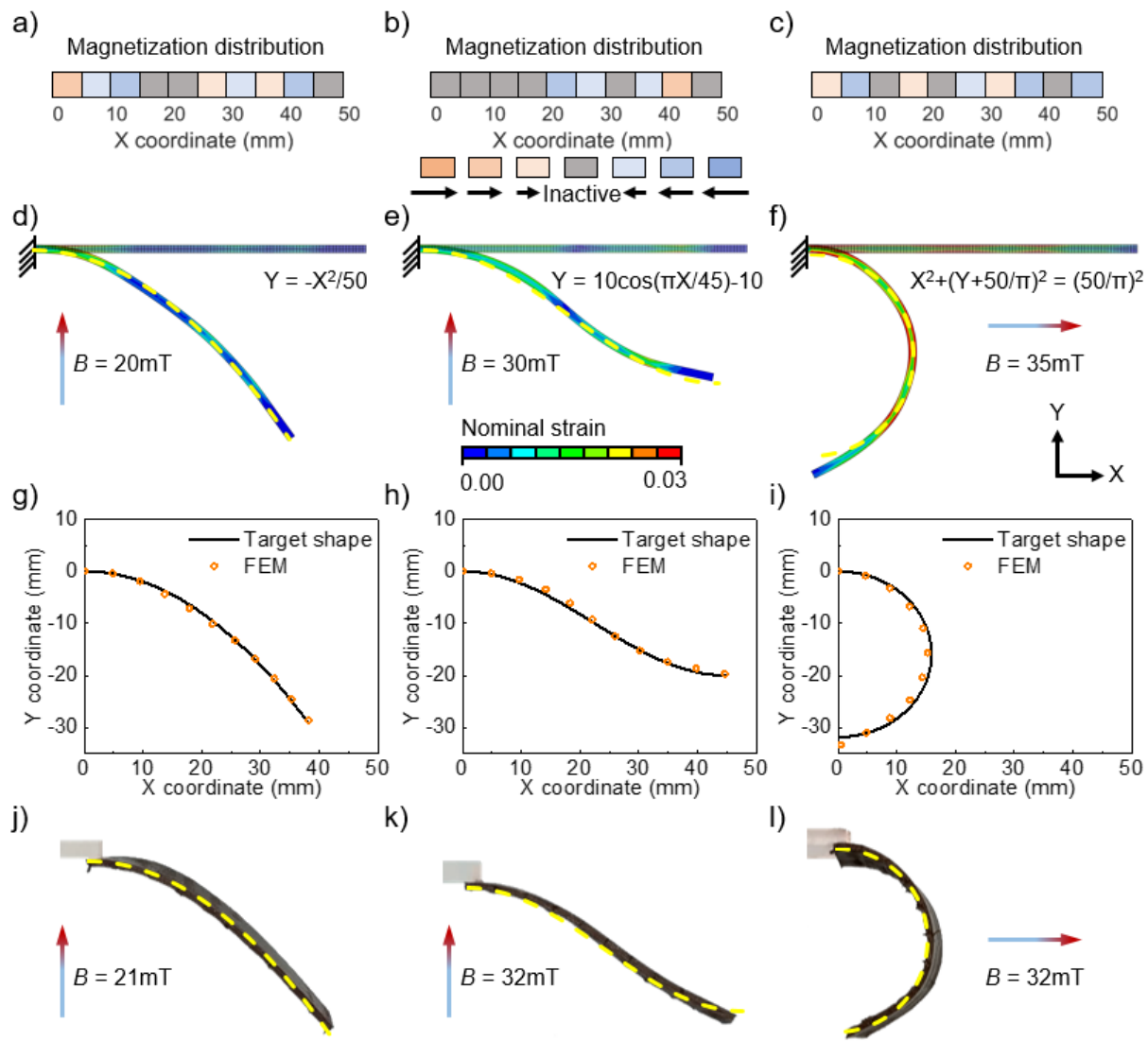

**Figure 5.** The EA-guided magnetization distribution designs and the voxel-encoding DIW printing for complicated morphing structures. a-c) Magnetization distributions by EA-guided design strategy with parabolic, cosine and half-circle curves as the target shapes. d-f) Deformations and maximum principal nominal strain distributions from FEM simulations under 20 mT, 30 mT and 35 mT with yellow dashed lines representing targeted deformations. g-i) Comparisons between the target shapes and FEM simulations. j-l) Deformations of the voxel-encoding DIW printed samples under 21 mT, 32 mT and 32 mT.

**A biomimetic soft crawling robot**

Through billions of years of evolution, nature develops efficient moving actions by dynamic shape morphing.[11] For example, during the crawling motion of an inchworm, the front and rear parts of the inchworm body remain contact with the ground to provide support and friction while the middle portion bends and recovers to provide the overall body motion (**Figure 6**a).[12] The neutral axis, which is denoted by the dashed black line in Figure 6a, transforms between a straight line and a highly curved line with a functional curvature distribution. The well-controlled curvature distribution of the body during morphing is the key to achieve the efficient motion. Utilizing the EA-guided design strategy with the voxel-encoding DIW printing, a biomimetic crawling motion can be realized.

Due to the symmetry, the shape of the right half of the inchworm body in Figure 6a is employed as the target shape. A 25 mm-long beam composed of three-layered 5 mm-long voxels is used for the FEM simulations and EA calculations. The symmetry and the roller support boundary conditions are applied to the left and right ends of the beam, respectively (Figure 6b). The target shape and the FEM simulation are shown in Figure 6c and are denoted by the solid black curve and the orange circles, respectively. The half model's magnetization distributions are applied to a full model with the magnetization distribution shown in Figure 6d. A 50 mm-long biomimetic crawling robot (Figure 6e) is then printed with ten three-layered 5 mm-long voxels. Upon applying a magnetic field, the middle portion of the robot bends up to a height of $h$ while the front and rear segments remain flat as predicted by the FEM simulation, mimicking the actual body curvature distribution of a crawling inchworm. When gradually removing the magnetic field, a smaller friction coefficient in the forward direction leads to a directional motion with a crawling distance of $\delta$. By applying a periodic magnetic field, the crawling robot achieves a biomimetic directional crawling motion as shown in Figure 6e and Video S3. Figure 6f shows the height $h$ (the solid blue line) and the crawling distance $\delta$ (the solid black line) as functions of the maximum amplitude of the applied magnetic field. The three inset images in Figure 6f show the deformations of the robot at 70 mT, 150 mT, and 300 mT, respectively. With a larger magnetic field, the middle segment of the robot bends higher, leading to a larger crawling distance when releasing the magnetic field. It should be noted that a larger magnetic field is required in the experiment than that in the simulation. This is because the friction between the robot and the substrate is ignored in the FEM simulations for the sake of simplicity. These results further demonstrate that the EA-guided magnetization distribution design with the voxel-encoding DIW printing facilitates achieving morphing structures with predetermined curvature distribution for functional applications.

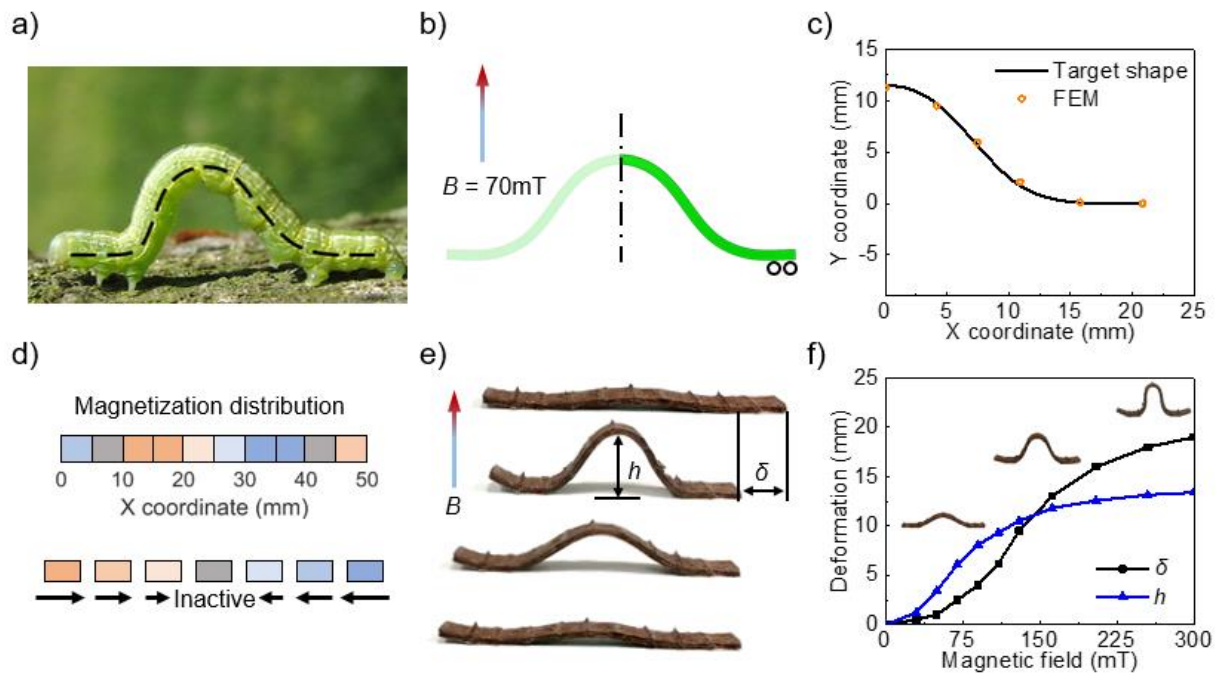

**Figure 6.** Biomimetic crawling motion with functional body curvature distribution via the EA-guided design of magnetization distribution. a) Crawling motion of an inchworm with the dashed black line representing the neutral axis. b) Deformation of the half model from the FEM simulation under 70 mT. c) Comparison between the target shape and the FEM simulation. d) Magnetization distribution of the full model. e) Crawling motion of the voxel-encoding DIW printed biomimetic robot. f) Height $h$ and crawling distance $\delta$ with respect to the applied magnetic field.

**A biomimetic soft walking robot**

To achieve complex and dexterous dynamic motions such as walking, running, and jumping, animals operate their different body parts in a synergistic way to maintain tractions and gain necessary speed.[13] For example, among different dog gaits, the trot is considered to be the most efficient one. It is a two-beat sequence where the diagonal opposite legs lift or strike the ground together to keep balance while moving forward. The legs on both sides work synergistically to move the body fast and efficiently, as shown in the schematics in **Figure 7**a.[14] To achieve this type of biomimetic motion, four-foot robots have been built, but heavily relying on the discrete control units such as multiple motors to generate motions of different body parts. This discrete control often requires a large effort in computing and communication between the control units, which as a consequence affect the operability and reliability of the robotic systems. To address these challenges, the synergistic operations of different body parts under a centralized control could potentially provide a solution. Here, we implement the synergistic motion by generating well-designed pace morphing of different body parts using our EA-guided voxel-encoding DIW printing of hmSAMs.

To obtain an effective dynamic motion by mimicking the dog trot, two curvature distributions, one parabolic and one sigmoidal, are taken as the target shapes simplified based on the geometries of the in-motion front and rear legs that are denoted by the blue curves in Figure 7a. To generate the biomimetic synergistic motion, we designed two target morphing shapes for the front and rear legs as illustrated in Figure 7b. The right and left legs are denoted by the blue solid curves and green dashed curves, respectively. The deformations from the FEM predictions guided by the EA are shown as blue circles. Here, 20 mm-long beams composed of six two-layered voxels are used, due to the consideration of the relatively low beam bending stiffness yet large enough magnetization design space ($5^6 = 15,625$). A 21 mm spacing is set between the front and rear legs to avoid interactions during walking. Based on the EA-guided design strategy of magnetization distributions, four functional hmSAM legs are DIW printed and glued onto the "body", which is made of a piece of thick paper. The morphing of the robot under a uniform 75 mT magnetic field is shown in Figure 7c. Here, the colored Teflon tapes are wrapped around the free end of the legs to reduce friction with green or blue corresponding to the left or right. The magnetization distributions of the front-right and rear-right legs are shown in Figure 7d. To lift and strike down each pair of the legs sequentially to mimic the two-beat trot gait, an alternating magnetic field is applied with right defined as positive direction (Figure 7e). The detailed gait of the biomimetic robot is shown in Figure 7f and Video S4. Initially, all four legs are on the ground. As a rightward (positive) magnetic field is applied, the front-right and rear-left legs lift at 0.2 s and strike the ground at 0.45 s while the other two legs deform but still remain on the ground. In Figure 7f, the rectangles with solid and dashed frames denote lifting and ground-touching, respectively. When a leftward (negative) magnetic field is applied, the motion switches to front-left and rear-right legs lifting at 0.8 s and striking down at 0.95 s. By alternating the magnetic field, the two pairs of diagonal-located legs move sequentially to form a two-beat trot gait. The hmSAM dog demonstrates the robustness of the EA-guided voxel-encoding DIW printing of hmSAMs in designing intelligent soft robotic systems with biomimetic synergistic motions. Different components of a soft robot can be predesigned according to actual shape morphing and dynamic motion needs, to achieve an assembly of highly functional parts with various curvature distributions.

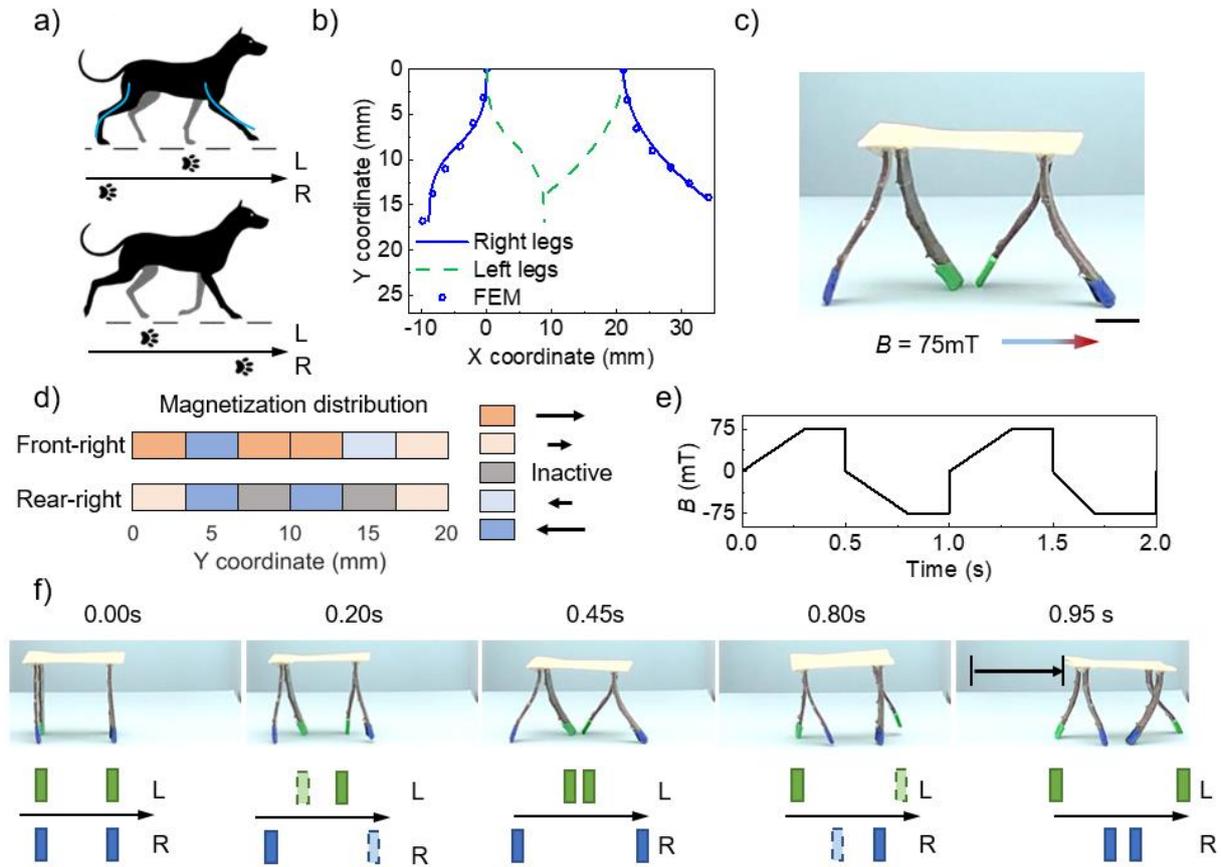

**Figure 7.** Biomimetic dog trot gait via the EA-guided design of the magnetization distribution. a) Schematics of dog trot motion. b) Designed target deformations and FEM predictions guided by EA design strategy. c) Actuation of a walking robot by voxel-encoding DIW printing under 75 mT. Scale bar: 5 mm. d) Magnetization distribution designs of two types of legs. e) Magnetic field profile applied for the biomimetic walking motion. f) Detailed demonstration of the biomimetic walking gait. The dashed frame means the corresponding leg is lifted and the solid frame means the leg touches the ground.

**Conclusion**

We propose a method of voxel-encoding DIW printing of hmSAMs to realize a wide range of magnetic property tunability for both magnetic density and magnetic direction distributions for the DIW printed hmSAMs. With the robust EA-guided design method, complex non-intuitive magnetization distributions to achieve complicated deformations with desired curvature distributions can be inversely designed. Utilizing the voxel-encoding DIW printing method and the EA-guided design strategy, two functional biomimetic soft robots are designed and fabricated, demonstrating the great potential of hmSAMs and our proposed approach to achieve rational magnetic actuations for practical applications. Moreover, the EA-guided design strategy could also be adopted to program the applied actuation field for functional actuation and motions with further complexity, which will be part of the future work. The combination of this inverse design technique and the remarkable magnetic property tunability in the magnetization of the hmSAMs via the proposed voxel-encoding DIW printing method can significantly broaden the application potentials of the hmSAMs.

**Methods**

*Preparation of the DIW printing*

The DIW ink of the hmSAMs contains two phases, hard-magnetic microparticles and silicone-based matrix. First, SE1700 base (Dow Corning Corp.) and Ecoflex 00-30 Part B (Smooth-on Inc.) in a volume ratio of 1:2 were mixed at 2000 rpm for 1 min by a centrifugal mixer (AR-100, Thinky Inc.). Then, 20 vol% NdFeB particles (77.5 vol% to SE1700 base, average particle size of 5 μm, Magnequench) were added to the above mixture and mixed at 2000 rpm for 2 min and defoamed at 2200 rpm for 3 min. SE1700 curing agent with 10 vol% to SE1700 base was added and mixed at 2000 rpm for 1 min. The well-mixed magnetic ink was transferred into 10cc syringe barrels (Nordson EFD), defoamed at 2200 rpm for 3 min to remove trapped air during transferring, and mixed at 2000 rpm for another 1 min. The ink was magnetized under a ~1.5 T impulse magnetic field generated by a customized impulse magnetizer. After magnetization, the syringe barrel was mounted to the custom-designed 3D printer (Aerotech), and SmoothFlow nozzle (Nordson EFD) with 410 μm inner diameter was used.

*Physical property characterization:*

The uniaxial tension tests were conducted on the mechanical tester DMA 850 (TA Instruments). DIW printed one-layered 20 mm by 4.8 mm thin film samples were stretched at a strain rate of 0.01 s$^{-1}$. Shear modulus was obtained by fitting the stress stretch curve with the Neo-Hookean model. The magnetizations of printed samples were measured by the vibrating sample magnetometer (VSM, 7707A Lake Shore Cryotronics, Inc.).

*Finite element analysis:*

To evaluate the magnetization distribution performance generated from the evolutionary algorithm, a user-defined element subroutine in commercial software ABAQUS 2019 (Dassault Systèmes) was used to predict the deformations.[5] The shear modulus, Poisson's ratio and magnetization were set to be 300 kPa, 0.495 and 70 kA m$^{-1}$ respectively, which were measured with the DIW printed one-layer 20 mm by 4.8 mm samples without joints.

**Supporting Information**

**Acknowledgements**
This work was supported in part by The Ohio State University Materials Research Seed Grant Program, funded by the Center for Emergent Materials, an NSF-MRSEC, grant DMR-1420451, the Center for Exploration of Novel Complex Materials, and the Institute for Materials Research. H.J.Q acknowledges the support of an AFOSR grant (FA9550-19-1-0151; Dr. B.-L. "Les" Lee, Program Manager).

References

[1]  a) G. Z. Lum, Z. Ye, X. Dong, H. Marvi, O. Erin, W. Hu, M. Sitti, *Proceedings of the National Academy of Sciences* **2016**, 113, E6007; b) W. Hu, G. Z. Lum, M. Mastrangeli, M. Sitti, *Nature* **2018**, 554, 81; c) Z. Ren, W. Hu, X. Dong, M. Sitti, *Nature communications* **2019**, 10, 1; d) T. Xu, J. Zhang, M. Salehizadeh, O. Onaizah, E. Diller, *Science Robotics* **2019**, 4, eaav4494; e) X. Du, H. Cui, T. Xu, C. Huang, Y. Wang, Q. Zhao, Y. Xu, X. Wu, *Advanced Functional Materials* **2020**, 1909202.
[2]  a) Y. Kim, H. Yuk, R. Zhao, S. A. Chester, X. Zhao, *Nature* **2018**, 558, 274; b) H. Gu, Q. Boehler, D. Ahmed, B. J. Nelson, *Science Robotics* **2019**, 4; c) S. Wu, Q. Ze, R. Zhang, N. Hu, Y. Cheng, F. Yang, R. Zhao, *ACS Applied Materials & Interfaces* **2019**, 11, 41649.
[3]  a) J. Cui, T.-Y. Huang, Z. Luo, P. Testa, H. Gu, X.-Z. Chen, B. J. Nelson, L. J. Heyderman, *Nature* **2019**, 575, 164; b) Q. Ze, X. Kuang, S. Wu, J. Wong, S. M. Montgomery, R. Zhang, J. M. Kovitz, F. Yang, H. J. Qi, R. Zhao, *Advanced Materials* **2019**.

[4]     a) S. Yim, M. Sitti, *IEEE Transactions on Robotics* **2011**, 28, 183; b) M. Pallapa, J. Yeow, *Journal of The Electrochemical Society* **2014**, 161, B3006; c) S. Jeon, A. K. Hoshiar, K. Kim, S. Lee, E. Kim, S. Lee, J.-y. Kim, B. J. Nelson, H.-J. Cha, B.-J. Yi, *Soft robotics* **2019**, 6, 54; d) Y. Kim, G. A. Parada, S. Liu, X. Zhao, *Science Robotics* **2019**, 4, eaax7329; e) J. Hwang, J.-y. Kim, H. Choi, *Intelligent Service Robotics* **2020**, 1.

[5]     R. Zhao, Y. Kim, S. A. Chester, P. Sharma, X. Zhao, *Journal of the Mechanics and Physics of Solids* **2019**, 124, 244.

[6]     a) W. M. van Rees, E. Vouga, L. Mahadevan, *Proceedings of the National Academy of Sciences* **2017**, 114, 11597; b) H. Aharoni, Y. Xia, X. Zhang, R. D. Kamien, S. Yang, *Proceedings of the National Academy of Sciences* **2018**, 115, 7206; c) T.-Y. Huang, H.-W. Huang, D. Jin, Q. Chen, J. Huang, L. Zhang, H. Duan, *Science Advances* **2020**, 6, eaav8219; d) E. Siéfert, E. Reyssat, J. Bico, B. Roman, *Nature materials* **2019**, 18, 24; e) J. W. Boley, W. M. van Rees, C. Lissandrello, M. N. Horenstein, R. L. Truby, A. Kotikian, J. A. Lewis, L. Mahadevan, *Proceedings of the National Academy of Sciences* **2019**, 116, 20856.

[7]     a) S. M. Secor, B. C. Jayne, A. F. Bennett, *Journal of experimental biology* **1992**, 163, 1; b) H. Marvi, C. Gong, N. Gravish, H. Astley, M. Travers, R. L. Hatton, J. R. Mendelson, H. Choset, D. L. Hu, D. I. Goldman, *Science* **2014**, 346, 224; c) H. C. Astley, C. Gong, J. Dai, M. Travers, M. M. Serrano, P. A. Vela, H. Choset, J. R. Mendelson, D. L. Hu, D. I. Goldman, *Proceedings of the National Academy of Sciences* **2015**, 112, 6200.

[8]     A. Patel, E. Boje, C. Fisher, L. Louis, E. Lane, *Biology open* **2016**, 5, 1072.

[9]     a) C. M. Hamel, D. J. Roach, K. N. Long, F. Demoly, M. L. Dunn, H. J. Qi, *Smart Materials and Structures* **2019**, 28, 065005; b) G. Sossou, F. Demoly, H. Belkebir, H. J. Qi, S. Gomes, G. Montavon, *Materials & Design* **2019**, 181, 108074.

[10]    D. Rainville, F.-A. Fortin, M.-A. Gardner, M. Parizeau, C. Gagné, presented at Proceedings of the 14th annual conference companion on Genetic and evolutionary computation **2012**.

[11]    a) A. S. Gladman, E. A. Matsumoto, R. G. Nuzzo, L. Mahadevan, J. A. Lewis, *Nature materials* **2016**, 15, 413; b) A. Nojoomi, H. Arslan, K. Lee, K. Yum, *Nature communications* **2018**, 9, 3705; c) H. Arslan, A. Nojoomi, J. Jeon, K. Yum, *Advanced Science* **2019**, 6, 1800703; d) S. A. Stamper, S. Sefati, N. J. Cowan, *Proceedings of the National Academy of Sciences* **2015**, 112, 5870; e) Y. Y. Xiao, Z. C. Jiang, X. Tong, Y. Zhao, *Advanced Materials* **2019**, 31, 1903452.

[12]    B. Matt, https://bugguide.net/node/view/101463, accessed: 12/10, 2019.

[13]    a) M. S. Triantafyllou, G. Triantafyllou, D. Yue, *Annual review of fluid mechanics* **2000**, 32, 33; b) S. Kim, M. Spenko, S. Trujillo, B. Heyneman, D. Santos, M. R. Cutkosky, *IEEE Transactions on robotics* **2008**, 24, 65; c) R. D. Maladen, Y. Ding, P. B. Umbanhowar, A. Kamor, D. I. Goldman, *Journal of The Royal Society Interface* **2011**, 8, 1332; d) C. Li, T. Zhang, D. I. Goldman, *science* **2013**, 339, 1408; e) S.-H. Song, M.-S. Kim, H. Rodrigue, J.-Y. Lee, J.-E. Shim, M.-C. Kim, W.-S. Chu, S.-H. Ahn, *Bioinspiration & biomimetics* **2016**, 11, 036010; f) Q. He, Z. Wang, Y. Wang, A. Minori, M. T. Tolley, S. Cai, *Science advances* **2019**, 5, eaax5746.

[14]    S. Cunnane, https://www.stephencunnane.com/3d-work, accessed: 12/10, 2019.